\documentclass[a4paper]{article}

\usepackage{arydshln}
\usepackage{amsmath}
\usepackage{mathtools}
\usepackage{amsfonts,amssymb}
\usepackage[compress]{cite}
\usepackage{graphicx}
\usepackage{mdwlist}
\usepackage{xfrac}
\usepackage{paralist}
\usepackage{algorithmic}
\usepackage{algorithm}
\usepackage{url}
\usepackage[font=footnotesize]{caption}

\newtheorem{lemma}{\bf Lemma}[section]
\newtheorem{theorem}[lemma]{\bf Theorem}
\newtheorem{remark}[lemma]{\bf Remark}

\newtheorem{assumption}[lemma]{\bf Assumption}

\renewcommand{\matrix}[2]{\left[\begin{array}{#1} #2 \end{array}\right] }

\newcommand{\jpfig}[4]{\begin{figure}[t] \centering \includegraphics[width=#1\linewidth]{#2} \caption{\label{#3}#4} \end{figure}}

\newcommand{\IEEEQED}{~\rule[-1pt]{5pt}{5pt}\par\medskip}
\newenvironment{IEEEproof}{{\noindent \bf Proof:\ }}{ \hfill \IEEEQED}

\DeclareMathOperator*{\trace}{trace}
\DeclareMathOperator*{\diag}{diag}

\floatname{algorithm}{Procedure}

\date{}

\begin{document}

\title{Complexity Reduction for Parameter-Dependent Linear Systems\thanks{The work was supported by the Swedish Research Council and the Knut and Alice Wallenberg Foundation. }}

\author{Farhad~Farokhi, Henrik~Sandberg,~and~Karl~H.~Johansson\thanks{The authors are with ACCESS Linnaeus Center, School of Electrical Engineering, KTH-Royal Institute of Technology, SE-100 44 Stockholm, Sweden. E-mails: \{farokhi,hsan,kallej\}@ee.kth.se }
}

\maketitle
\begin{abstract} We present a complexity reduction algorithm for a family of parameter-dependent linear systems when the system parameters belong to a compact semi-algebraic set. This algorithm potentially describes the underlying dynamical system with fewer parameters or state variables. To do so, it minimizes the distance (i.e., $H_\infty$-norm of the difference) between the original system and its reduced version. We present a sub-optimal solution to this problem using sum-of-squares optimization methods. We present the results for both continuous-time and discrete-time systems. Lastly, we illustrate the applicability of our proposed algorithm on numerical examples.
\end{abstract}
\vspace{-.1in}
\section{Introduction} \label{sec:intro}
Large-scale systems are often composed of several interacting subsystems described by local parameters that need to be identified when designing model-based control laws. The parameters are typically a function of the working points of the subsystems and their physical properties. Hence, they vary over time based on the operation mode. In practice, we like to develop a family of controllers that only depend on a few of the system parameters, such that we do not need to adjust the whole controller whenever a parameter changes in the system. In addition, we might want to study the relative importance of the system parameters. Gain scheduling and supervisory control are examples of parameter-dependent controllers~\cite{packard1994gain,rugh2000research,shamma1992gain,leith2000survey,nichols1993gain,morse1996supervisory,scorletti1998improved}.
However, these design methods implicitly assume in most cases that the overall controller has access to the entire set of model parameters. This assumption might not be realistic in many practical cases (see~\cite{Farokhi-thesis2012} and references therein for a detailed discussion). Hence, we are interested in introducing a complexity reduction algorithm to effectively remove some of the system parameters or to decrease its order while preserving the input-output transfer function to some extent. Doing so, we can then simplify the control design procedure or satisfy the requirements described above on model parameter dependencies. The problem of model reduction for parameter-dependent linear systems has been studied extensively~\cite{Li5373988,beck1996model,halevi1997parameter,dullerud2000course,beck2001formal}. For instance, the authors in~\cite{beck1996model} used a multidimensional system formulation and introduced a generalization of controllability and observability Gramians using a pair of linear matrix inequalities (LMIs). Using these generalized Gramians, they performed balanced truncation to extract the reduced system. They also calculated an upper bound for the error of this truncation. However, the reduced order system presented in~\cite{beck1996model} is not optimal since the introduced upper bound for the truncation error is not necessarily tight. In this paper, we introduce a near-optimal numerical procedure for extracting these reduced system.

Specifically, we are interested in minimizing the $H_\infty$-norm of the difference of the original system transfer functions and its reduced version over a compact semi-algebraic set of system parameters. Using the bounded real lemma~\cite{zhou1998essentials}, we transform this problem to a parameter-dependent feasibility-checking bilinear matrix inequality (BMI). We use the method of alternating LMIs~\cite{springerlink:10.1007/BF01099648} to transform this parameter-dependent BMI into a string of LMIs. Then, we use the method introduced in~\cite{springerlink:10.1007/s10107-005-0684-2} to solve these parameter-dependent LMIs by means of sum-of-squares optimization. This algorithm results in a sub-optimal solution because~(1)~when using the method of alternating LMIs, we cannot guarantee the convergence of the proposed algorithm (i.e., there exists always a BMI such that you cannot check its feasibility using the method of alternating LMIs~\cite{springerlink:10.1007/BF01099648}), and~(2)~when using sum-of-squares optimization for solving the parameter-dependent LMIs, the lack of convergence to a solution does not imply the infeasibility of the original problem (since a sum-of-squares matrix is indeed a positive-definite polynomial matrix but not the other way around)~\cite{springerlink:10.1007/s10107-005-0684-2}. Due to relying on sum-of-square optimization, the proposed algorithm does not scale well with the system dimension and the number of parameters. However, we might be able to exploit sparsity patterns or symmetry structures in future to develop better numerical algorithms~\cite{6161526}. Despite these inefficiencies, we observe that the proposed algorithm is fairly strong in solving the proposed numerical examples in Section~\ref{sec:illustrativeexample}.

Recently, there have been many studies on using sum-of-squares optimization methods in control design~\cite{papachristodoulou2005analysis,papachristodoulou2002construction,Lavaei2010110,parrilo2003semidefinite}.
For instance, the problem of finding a polynomial Lyapunov function for nonlinear systems was considered in~\cite{papachristodoulou2005analysis,papachristodoulou2002construction}. The problem of optimal linear quadratic control design for parameter-dependent discrete-time systems was discussed in~\cite{Lavaei2010110}. However, to the best of our knowledge, no attention has been paid to complexity reduction using sum-of-square optimization.

The rest of the paper is organized as follows. In Section~\ref{sec:problemformulation}, we present the mathematical problem formulation. We introduce our complexity reduction algorithm and prove its suboptimality in Section~\ref{sec:mainresults}. We illustrate the applicability of the proposed algorithm on two numerical examples and compare their results with available methods in Section~\ref{sec:illustrativeexample}. Finally, we conclude the paper in Section~\ref{sec:conclusion}.

\subsection{Notation}
The sets of integer, natural, real, and complex numbers are denoted respectively by $\mathbb{Z}$, $\mathbb{N}$, $\mathbb{R}$, and $\mathbb{C}$. For any integer number $n\in\mathbb{Z}$ and any real number $x\in\mathbb{R}$, we define the notations $\mathbb{Z}_{> (\geq) n}=\{m\in\mathbb{Z}\;|\; m> (\geq) n\}$ and $\mathbb{R}_{> (\geq) x}=\{y\in\mathbb{R}\;|\; y> (\geq) x\}$, respectively. All other sets are denoted by calligraphic letters such as $\mathcal{A}$ and $\mathcal{B}$.

Matrices are denoted by capital roman letters such as $A$ and $B$. $A > (\geq) 0$ means that the symmetric matrix $A\in \mathbb{R}^{n\times n}$ is positive definite (positive semidefinite) and $A > (\geq) B$ implies that $A-B > (\geq) 0$.

The ring of polynomials with coefficients in $\mathbb{R}$ is denoted by $\mathbb{R}[\alpha]$, where $\alpha$ is the vector of variables. For any given $n,m\in\mathbb{N}$, a polynomial matrix $X(\alpha) \in \mathbb{R}[\alpha]^{n\times m}$ is a matrix whose entries are polynomials in $\mathbb{R}[\alpha]$, that is, $x_{ij}(\alpha)\in\mathbb{R}[\alpha]$ for $1\leq i\leq n$ and $1\leq j\leq m$. For any given $n\in\mathbb{N}$, a matrix polynomial $X(\alpha)\in\mathbb{R}[\alpha]^{n\times n}$ is positive definite (positive semidefinite) if for each $\alpha\in\mathcal{A}$, the matrix $X(\alpha)\in\mathbb{R}^{n\times n}$ is positive definite (positive semidefinite), where the set $\mathcal{A}$ will be defined in the text.

For any given $n\in\mathbb{N}$, a matrix polynomial $X(\alpha)\in\mathbb{R}[\alpha]^{n \times n}$ is a sum-of-square matrix, denoted by $X(\alpha)\succeq 0$, if there exits a matrix polynomial $Y(\alpha)\in\mathbb{R}[\alpha]^{n \times n}$ such that $X(\alpha)=Y(\alpha)^\top Y(\alpha)$. We introduce the notation $\mathcal{S}[\alpha]^n=\{X(\alpha)\in\mathbb{R}[\alpha]^{n \times n}\;|\;X(\alpha)\succeq 0\}$ to capture the set of all sum-of-square matrices. When $n=1$, we use $\mathcal{S}[\alpha]$ instead of $\mathcal{S}[\alpha]^1$.

\section{Problem Formulation} \label{sec:problemformulation}
In this section, we present the mathematical formulation of the complexity reduction problem introduced in Section~\ref{sec:intro} for both continuous-time and discrete-time parameter-dependent linear systems.

\subsection{Continuous-Time Systems}
Consider a parameter-dependent continuous-time linear dynamical system described by
\begin{equation} \label{enq:originalsystem}
\begin{split}
\begin{array}{rl} G(s;\alpha): & \hspace{-.1in}\left\{\hspace{-.06in}\begin{array}{l} \dot{x}(t)=A(\alpha)x(t)+B(\alpha)u(t),\\
y(t)=C(\alpha)x(t)+D(\alpha)u(t), \end{array} \right. \end{array}
\end{split}
\end{equation}
where $x(t)\in\mathbb{R}^n$ is the state vector, $u(t)\in\mathbb{R}^m$ is the control input, $y(t)\in\mathbb{R}^o$ is the system output, and $\alpha\in\mathbb{R}^p$ is the parameter vector. Note that in~(\ref{enq:originalsystem}), we use the notation
$$
G(s;\alpha)=C(\alpha)(sI-A(\alpha))^{-1}B(\alpha)+D(\alpha).
$$

Throughout this paper, we assume that $\alpha\in\mathcal{A}\subset\mathbb{R}^p$, where $\mathcal{A}$ is defined to be the set of eligible parameters. We are interested in extracting a reduced parameter-dependent continuous-time linear system described by
\begin{equation} \label{enq:reducedsystem}
\begin{split}
\begin{array}{rl} G'(s;\alpha'): & \hspace{-.1in}\left\{\hspace{-.06in}\begin{array}{l} \dot{x}'(t)=A'(\alpha')x'(t)+B'(\alpha')u(t),\\
y'(t)=C'(\alpha')x'(t)+D'(\alpha')u(t), \end{array} \right. \end{array}
\end{split}
\end{equation}
where $x'(t)\in\mathbb{R}^{n'}$ is the reduced system state vector, $y'(t)\in\mathbb{R}^{o}$ is its output, and $\alpha'\in\mathcal{A}'\subset\mathbb{R}^{p'}$ is the reduced parameter vector. Note that the output vector dimension stays the same. We define the reduced set of eligible parameters as
$$
\mathcal{A}'=\left\{\alpha'\in\mathbb{R}^{p'}\;\big|\; \exists \xi\in\mathbb{R}^{p-p'} : \matrix{c}{\alpha' \\ \xi }\in\mathcal{A} \right\}.
$$

\begin{remark} We name this procedure as complexity reduction because we can potentially reduce the number of the parameters with which the system is described (since, by definition, we assume $p'\leq p$). In addition, by choosing $n'\leq n$, we may also reduce the system order.
\end{remark}

Throughout this paper, we make the following assumption concerning the model matrices:
\begin{assumption} \label{assumption1} The model matrices in~(\ref{enq:originalsystem}) and~(\ref{enq:reducedsystem}) are polynomials in terms of the system parameters~$\alpha$ and~$\alpha'$, that is, $A(\alpha)\in\mathbb{R}[\alpha]^{n\times n}$, $B(\alpha)\in\mathbb{R}[\alpha]^{n\times m}$, $C(\alpha)\in\mathbb{R}[\alpha]^{o\times n}$, $D(\alpha)\in\mathbb{R}[\alpha]^{o\times m}$, $A'(\alpha')\in\mathbb{R}[\alpha']^{n'\times n'}$, $B'(\alpha')\in\mathbb{R}[\alpha']^{n'\times m}$, $C'(\alpha')\in\mathbb{R}[\alpha']^{o\times n'}$, and $D'(\alpha')\in\mathbb{R}[\alpha']^{o\times m}$.
\end{assumption}

We are interested in finding $G'(s;\alpha')$ to minimize the distance between the systems in~(\ref{enq:originalsystem}) and~(\ref{enq:reducedsystem}):
\begin{equation} \label{eqn:optimization1}
\inf_{G'(s;\alpha')} \sup_{\alpha\in\mathcal{A}} \left\|G(s;\alpha)-G'(s;T(\alpha)) \right\|_\infty,
\end{equation}
where the projection $T:\mathbb{R}^{p} \rightarrow \mathbb{R}^{p'}$ is defined as
$T(x)=[x_1 \cdots x_{p'}]^\top$ for all $x\in\mathbb{R}^{p}$. The optimization problem in~(\ref{eqn:optimization1}) is to be solved subject to the reduced system state-space description~(\ref{enq:reducedsystem}) and the fact that the model matrices are polynomial matrices in $\alpha$ and $\alpha'$ (Assumption~\ref{assumption1}).

\begin{remark} If we are ultimately interested in designing a controller using the reduced system, we should solve the optimization problem
\begin{equation*}
\inf_{G'(s;\alpha')} \sup_{\alpha\in\mathcal{A}} \left\|G(s;\alpha)^{-1}(G(s;\alpha)-G'(s;T(\alpha)) ) \right\|_\infty,
\end{equation*}
 see~\cite{obinata2000model}. In the case that $G(s;\alpha)$ does not vary much over the set of eligible parameter~$\mathcal{A}$, we can instead solve the optimization problem
\begin{equation} \label{eqn:newoptimization}
\inf_{G'(s;\alpha')} \sup_{\alpha\in\mathcal{A}} \left\|G(s;\beta)^{-1}(G(s;\alpha)-G'(s;T(\alpha)) ) \right\|_\infty,
\end{equation}
for some fixed $\beta\in\mathcal{A}$. As our developed algorithm would not change much for solving~(\ref{eqn:newoptimization}) instead of~(\ref{eqn:optimization1}), we would only focus on solving~(\ref{eqn:optimization1}) in this paper.
\end{remark}

\subsection{Discrete-Time Systems}
Consider a parameter-dependent discrete-time linear time-invariant system
\begin{equation} \label{enq:discrete:originalsystem}
\begin{split}
\begin{array}{rl} G(\mathfrak{z};\alpha): & \hspace{-.1in}\left\{\hspace{-.06in}\begin{array}{rcl} x(k+1)&\hspace{-.1in}=&\hspace{-.1in}A(\alpha)x(k)+B(\alpha)u(k),\\
y(k)&\hspace{-.10in}=&\hspace{-.10in}C(\alpha)x(k)+D(\alpha)u(k), \end{array} \right. \end{array}
\end{split}
\end{equation}
where, similar to the previous subsection, $x(k)\in\mathbb{R}^n$ is the state vector, $u(k)\in\mathbb{R}^m$ is the control input, $y(k)\in\mathbb{R}^o$ is the system output, and $\alpha\in\mathbb{R}^p$ is the parameter vector. In~(\ref{enq:discrete:originalsystem}), we use the notation
$$
G(\mathfrak{z};\alpha)=C(\alpha)(\mathfrak{z}I-A(\alpha))^{-1}B(\alpha)+D(\alpha).
$$
We define the reduced system as
\begin{equation} \label{enq:discrete:reducedsystem}
\begin{split}
\begin{array}{rl} \hspace{-.13in} G'(\mathfrak{z};\alpha'): & \hspace{-.1in}\left\{\hspace{-.06in}\begin{array}{rcl} x'(k+1)&\hspace{-.1in}=&\hspace{-.1in}A'(\alpha')x'(k)+B'(\alpha')u(k),\\
y'(k)&\hspace{-.1in}=&\hspace{-.1in}C'(\alpha')x'(k)+D'(\alpha')u(k), \end{array} \right. \end{array}
\end{split}
\end{equation}
where $x'(t)\in\mathbb{R}^{n'}$ is the reduced system state vector, $y'(t)\in\mathbb{R}^{o}$ is its output, and $\alpha'\in\mathcal{A}'\subset\mathbb{R}^{p'}$ is the reduced parameter vector. For these parameter-dependent discrete-time systems, we are interested in solving the optimization problem
\begin{equation} \label{eqn:optimization2}
\inf_{G'(\mathfrak{z};\alpha')} \sup_{\alpha\in\mathcal{A}}  \left\|G(\mathfrak{z};\alpha)-G'(\mathfrak{z};T(\alpha)) \right\|_\infty,
\end{equation}
subject to the reduced system state-space description in~(\ref{enq:reducedsystem}) and Assumption~\ref{assumption1}. In the next section, we present solutions to the optimization problems~(\ref{eqn:optimization1}) and~(\ref{eqn:optimization2}).

\section{Main Results} \label{sec:mainresults}
In this section, we rewrite the optimization problems as parameter-dependent feasibility-checking BMIs. We use the method of alternating LMIs to transform this parameter-dependent BMI into a string of LMIs, which we then solve using sum-of-squares optimization methods. First, we present the solution for continuous-time systems.

\begin{algorithm*}[t!]
\caption{Extracting the sub-optimal reduced system $G'(s;\alpha')$.}
\label{alg:1}
\begin{algorithmic}
\REQUIRE $\delta\in\mathbb{R}_{\geq 0}$, $\epsilon \in\mathbb{R}_{\geq 0}$, $\gamma\in\mathbb{R}_{\geq 0}$, $n'\in\mathbb{N}$, $p'\in\mathbb{N}$, $p'\in\mathbb{N}$, $d_P\in\mathbb{N}$, $d_A\in\mathbb{N}$, $d_B\in\mathbb{N}$, $d_C\in\mathbb{N}$, $d_D\in\mathbb{N}$, and $d_{Q_\ell}\in\mathbb{N}$ for all $q\leq \ell\leq L$.
\ENSURE $A'(\alpha')\in\mathbb{R}[\alpha']^{n'\times n'}$, $B'(\alpha')\in\mathbb{R}[\alpha']^{n'\times m}$, $C'(\alpha')\in\mathbb{R}[\alpha']^{o\times n'}$, $D'(\alpha')\in\mathbb{R}[\alpha']^{o\times m}$, and $P(\alpha)\in\mathcal{S}[\alpha]^{n+n'}$.
\STATE \hspace{-.13in}\textbf{Initialization:} Pick $A'(\alpha')\in\mathbb{R}[\alpha']^{n'\times n'}$, $B'(\alpha')\in\mathbb{R}[\alpha']^{n'\times m}$, $C'(\alpha')\in\mathbb{R}[\alpha']^{o\times n'}$, and $D'(\alpha')\in\mathbb{R}[\alpha']^{o\times m}$ such that $\deg(A'(\alpha'))=d_A$, $\deg(B'(\alpha'))=d_B$, $\deg(C'(\alpha'))=d_C$, and $\deg(D'(\alpha'))=d_D$, respectively. Also, set $P_{\mathrm{old}}(\alpha)=0$.
\FOR{$i=1,2,\dots$}
\STATE Find polynomial matrix $P(\alpha)\in\mathcal{S}[\alpha]^{n+n'}$ with $\deg(P(\alpha))=d_P$ and polynomial matrices $Q_\ell(\alpha)\in\mathcal{S}[\alpha]^{n+n'+m+o}$ with $\deg(Q_\ell(\alpha))=d_{Q_\ell}$ for all $1\leq \ell \leq L$, such that
\begin{small}
\begin{equation*} \tag{P.1} \label{eqn:P1}
\begin{split}
&\matrix{ccc}{\tilde{A}(\alpha)^\top P(\alpha)+P(\alpha)\tilde{A}(\alpha) & * & * \\
\tilde{B}(\alpha)^\top P(\alpha) & -I & * \\ \tilde{C}(\alpha) & \tilde{D}(\alpha) & -\gamma^2 I } + Q_0+\sum_{\ell=1}^L Q_\ell(\alpha) q_\ell(\alpha)= 0.
\end{split}
\end{equation*}
\end{small}
\STATE Find polynomial matrices $Q_\ell(\alpha)\in\mathcal{S}[\alpha]^{n+n'+m+o}$ with $\deg(Q_\ell(\alpha))=d_{Q_\ell}$ for all $1\leq \ell \leq L$ and model matrices $A'(\alpha')\in\mathbb{R}[\alpha']^{n'\times n'}$, $B'(\alpha')\in\mathbb{R}[\alpha']^{n'\times m}$, $C'(\alpha')\in\mathbb{R}[\alpha']^{o\times n'}$, and $D'(\alpha')\in\mathbb{R}[\alpha']^{o\times m}$ with respectively $\deg(A'(\alpha'))=d_A$, $\deg(B'(\alpha'))=d_B$, $\deg(C'(\alpha'))=d_C$, and $\deg(D'(\alpha'))=d_D$ such that
\begin{small}
\begin{equation*} \tag{P.2} \label{eqn:P2}
\begin{split}
&\matrix{ccc}{\tilde{A}(\alpha)^\top P(\alpha)+P(\alpha)\tilde{A}(\alpha) & * & * \\
\tilde{B}(\alpha)^\top P(\alpha) & -I & * \\ \tilde{C}(\alpha) & \tilde{D}(\alpha) & -\gamma^2 I } + Q_0+\sum_{\ell=1}^L Q_\ell(\alpha) q_\ell(\alpha)= 0.
\end{split}
\end{equation*}
\end{small}
\IF{ $\max_{\alpha\in\mathcal{A}}\|P(\alpha)-P_{\mathrm{old}}(\alpha)\|\leq \delta$ }
\STATE \textbf{break}
\ELSE
\STATE $P_{\mathrm{old}}(\alpha)\leftarrow P(\alpha)$
\ENDIF
\ENDFOR
\end{algorithmic}
\end{algorithm*}

\subsection{Complexity Reduction for Continuous-Time Systems}
Before stating the results, let us define the augmented system as
\begin{equation} \label{eqn:augmentedsystem}
\begin{split}
\matrix{c}{\dot{x}(t)\\ \dot{x}'(t)}=\tilde{A}(\alpha) \matrix{c}{x(t)\\ x'(t)}+\tilde{B}(\alpha)u(t),
\\ y(t)-y'(t)=\tilde{C}(\alpha) \matrix{c}{x(t)\\ x'(t)}+\tilde{D}(\alpha) u(t).
\end{split}
\end{equation}
where
\begin{equation} \label{eqn:augmentedsystem:A&B}
\tilde{A}(\alpha)=\matrix{cc}{A(\alpha) & 0 \\ 0 & A'(\alpha')}, \hspace{.1in}\tilde{B}(\alpha)=\matrix{c}{B(\alpha) \\ B'(\alpha')},
\end{equation}
and
\begin{equation} \label{eqn:augmentedsystem:C&D}
\tilde{C}(\alpha)=\matrix{cc}{C(\alpha) & -C'(\alpha')}, \hspace{.1in}\tilde{D}(\alpha)=D(\alpha)-D'(\alpha').
\end{equation}
Now, we are ready to present the first result of the paper. The next lemma transforms the $H_\infty$-optimization problem in~(\ref{eqn:optimization1}) into a parameter-dependent BMI.

\begin{lemma} \label{lemma:LMI} For a fixed $\alpha\in\mathcal{A}$ and $G'(s;\alpha')$, we have
$\|G(s;\alpha)-G'(s;\alpha') \|_\infty \leq \gamma$
if and only if there exists $P(\alpha)=P(\alpha)^\top\in\mathbb{R}^{(n+n')\times (n+n')}$ such that
$P(\alpha)\geq 0$ and
\begin{equation} \label{eqn:LMI}
\matrix{ccc}{\tilde{A}(\alpha)^\top P(\alpha)+P(\alpha)\tilde{A}(\alpha) & * & * \\
\tilde{B}(\alpha)^\top P(\alpha) & -I & * \\ \tilde{C}(\alpha) & \tilde{D}(\alpha) & -\gamma^2 I } \leq 0,
\end{equation}
for all $\alpha\in\mathcal{A}$.
\end{lemma}

\begin{IEEEproof} The proof follows from Bounded Real Lemma~\cite{zhou1998essentials} on the augmented system~(\ref{eqn:augmentedsystem}). Note that after fixing $\alpha\in\mathcal{A}$ and $G'(s;\alpha')$, the augmented system is simply a linear time-invariant system.
\end{IEEEproof}

Note that Lemma~\ref{lemma:LMI} does not guarantee that $P(\alpha)$ is a matrix polynomial in~$\alpha$. In the next lemma, we show that this is indeed the case using the results in~\cite{Bliman2004165}.

\begin{lemma} \label{tho:1} Let $\mathcal{A}$ be a compact subset of $\mathbb{R}^p$. Then, for a fixed $G'(s;\alpha')$, $\|G(s;\alpha)-G'(s;\alpha') \|_\infty \leq \gamma$ for all $\alpha\in\mathcal{A}$ if and only if there exists a positive definite polynomial matrix $P(\alpha)\in\mathbb{R}[\alpha]^{(n+n')\times (n+n')}$ such that the inequality in~(\ref{eqn:LMI}) is satisfied for all $\alpha\in\mathcal{A}$.
\end{lemma}

\begin{IEEEproof} Follows from Theorem~1 in~\cite{Bliman2004165} together with Lemma~\ref{lemma:LMI} above. \end{IEEEproof}

\begin{remark} To check the condition in Lemma~\ref{tho:1}, first, we should pick an integer $d_P\in\mathbb{N}$ and search over the set of all positive definite polynomial matrices $P(\alpha)\in\mathbb{R}[\alpha]^{(n+n')\times (n+n')}$ such that $\deg(P(\alpha))\leq d_P$, in order to find a feasible solution to the inequality in~(\ref{eqn:LMI}) for all $\alpha\in\mathcal{A}$. Now, since the degree of $P(\alpha)$ is not known in advance, we have to start from an initial value (possibly estimated based on intuition from the physical nature of the problem) and keep increasing $d_P$ until we reach a feasible solution, which exists if the distance $\|G(s;\alpha)-G'(s;\alpha') \|_\infty$ is less than $\gamma$. Therefore, we should also start with large values for $\gamma$ (to ensure the existence of a feasible solution) and then, decrease $\gamma$ accordingly (for instance, using the bisection method~\cite{atkinson1989introduction}).
\end{remark}

In the next theorem, we use sum-of-squares optimization to rewrite the inequality in~(\ref{eqn:LMI}) as a sum-of-square feasibility problem which we use later to develop our numerical algorithm.

\begin{theorem} \label{tho:2} Assume that the compact set $\mathcal{A}$ can be characterized as
$$
\mathcal{A}=\left\{\alpha\in\mathbb{R}^{p} \;|\; q_\ell(\alpha)\geq 0, \forall 1\leq \ell \leq L  \right\},
$$
where $q_\ell\in\mathbb{R}[\alpha]$ for all $1\leq \ell \leq L$. Furthermore, assume that there exist $w_\ell\in\mathcal{S}[\alpha]$ for all $0\leq \ell \leq L$, such that $\{\alpha\in\mathbb{R}^{p}\;\big|\; w_0(\alpha)+\sum_{\ell=1}^L q_\ell(\alpha)w_\ell(\alpha)\geq 0 \}$ is a compact set. Then, for a fixed $G'(s;\alpha')$, we have $\|G(s;\alpha)-G'(s;\alpha') \|_\infty \leq \gamma$ for all $\alpha\in\mathcal{A}$ if there exist a constant $\epsilon\in\mathbb{R}_{>0}$, polynomial matrices $P(\alpha)\in\mathcal{S}[\alpha]^{n+n'}$, and $Q_\ell(\alpha)\in\mathcal{S}[\alpha]^{n+n'+m+o}$ for all $1\leq \ell\leq L$, such that
\begin{equation} \label{eqn:polyEqu}
\begin{split}
&\matrix{ccc}{\tilde{A}(\alpha)^\top P(\alpha)+P(\alpha)\tilde{A}(\alpha) & * & * \\
\tilde{B}(\alpha)^\top P(\alpha) & -I & * \\ \tilde{C}(\alpha) & \tilde{D}(\alpha) & -\gamma^2 I } +\epsilon I+ Q_0+\sum_{\ell=1}^L Q_\ell(\alpha) q_\ell(\alpha)= 0.
\end{split}
\end{equation}
\end{theorem}

\begin{IEEEproof} The proof follows from Theorem~2 in~\cite{springerlink:10.1007/s10107-005-0684-2} together with Lemma~\ref{tho:1} above.
\end{IEEEproof}

\begin{remark} To check the condition in Theorem~\ref{tho:2}, we should pick the polynomial degree $d_P\in\mathbb{N}$ and search over the set of all sum-of-square polynomial matrices $P(\alpha)\in\mathcal{S}[\alpha]^{(n+n')\times (n+n')}$ such that $\deg(P(\alpha))\leq d_P$, in order to solve the polynomial equation in~(\ref{eqn:polyEqu}). This search is easy to perform since the underlying problem is convex (due to the restriction to the set of sum-of-square polynomial matrices) and can be readily solved using available LMI solvers. Unfortunately, if we cannot find any solution to this problem for a given degree $d_P$, we cannot deduce that our problem does not admit a solution for this given degree~$d_P$, since Theorem~\ref{tho:2} is only a sufficiency result. We can only hope to find a solution by increasing the polynomial degree~$d_P$.
\end{remark}

Note that so far, we assumed that the model matrices $A'(\alpha')$, $B'(\alpha')$, $C'(\alpha')$, and $D'(\alpha')$ are given since otherwise, Theorem~\ref{tho:2} would result in nonlinear equations in terms of unknown polynomial coefficients. We propose Procedure~\ref{alg:1} for finding matrices $A'(\alpha')$, $B'(\alpha')$, $C'(\alpha')$, and $D'(\alpha')$ based on the method of alternating LMIs for solving BMIs~\cite{springerlink:10.1007/BF01099648}.

\begin{remark} This method does not guarantee convergence for the proposed algorithm because there exists always at least one BMI which we cannot check its feasibility using the method of alternating LMIs~\cite{springerlink:10.1007/BF01099648}.
\end{remark}

\subsection{Complexity Reduction for Discrete-Time Systems}
In the next theorem, we present a result which is a counterpart to Theorem~\ref{tho:2} for discrete-time systems.

\begin{theorem} \label{tho:2} Assume that the compact set $\mathcal{A}$ can be characterized as
$$
\mathcal{A}=\left\{\alpha\in\mathbb{R}^{p} \;|\; q_\ell(\alpha)\geq 0, \forall 1\leq \ell \leq L  \right\},
$$
where $q_\ell\in\mathbb{R}[\alpha]$ for all $1\leq \ell \leq L$. Furthermore, assume that there exist $w_i\in\mathcal{S}[\alpha]$ for all $0\leq \ell \leq L$, such that $\{\alpha\in\mathbb{R}^{p}\;\big|\; w_0(\alpha)+\sum_{\ell=1}^L q_\ell(\alpha)w_\ell(\alpha)\geq 0 \}$ is a compact set. Then, for a fixed $G'(\mathfrak{z};\alpha')$, we have $\|G(\mathfrak{z};\alpha)-G'(\mathfrak{z};\alpha') \|_\infty \leq \gamma$ for all $\alpha\in\mathcal{A}$ if there exist a constant $\epsilon\in\mathbb{R}_{>0}$, polynomial matrices $P(\alpha)\in\mathcal{S}[\alpha]^{n+n'}$, and $Q_\ell(\alpha)\in\mathcal{S}[\alpha]^{2(n+n')+m+o}$ for all $1\leq \ell\leq L$, such that
\begin{equation} \label{eqn:polyEqu}
\begin{split}
&\matrix{cccc}{P(\alpha) & * & * & * \\ P(\alpha) \tilde{A}(\alpha)^\top & P(\alpha) & * & * \\
\tilde{B}(\alpha)^\top & 0 & I & * \\ 0 & \tilde{C}(\alpha)P(\alpha) & \tilde{D}(\alpha) & \gamma^2 I} -\epsilon I- Q_0-\sum_{\ell=1}^L Q_\ell(\alpha) q_\ell(\alpha)= 0.
\end{split}
\end{equation}
\end{theorem}

\begin{IEEEproof} The proof follows the same reasoning as in Lemmas~\ref{lemma:LMI}--\ref{tho:1} and Theorem~\ref{tho:2}.
\end{IEEEproof}

We can construct a similar procedure for discrete-time systems as in Procedure~\ref{alg:1} by changing the nonlinear equations in~(\ref{eqn:P1})--(\ref{eqn:P2}) with the nonlinear equation in~(\ref{eqn:polyEqu}) to calculate the reduced discrete-time system.

\section{Illustrative Example} \label{sec:illustrativeexample}
In this subsection, we illustrate the applicability of the developed procedure on two numerical examples. The first numerical example is a parameter-dependent discrete-time linear systems. We use this example to compare our developed algorithm with the method described in~\cite{beck1996model}. The second example is a parameter-dependent continuous-time linear system motivated by controlling power systems. To implement Procedure~\ref{alg:1}, we used SOSTOOLS which is a free MATLAB\textsuperscript{\textregistered} toolbox for formulating and solving sum-of-squares optimizations~\cite{sostools}.

\subsection{Discrete-Time Systems}
Consider the parameter-dependent discrete-time linear system described by
$$
\matrix{c}{ \hspace{-.05in} x_1(k+1) \hspace{-.05in} \\ \hspace{-.05in} x_2(k+1) \hspace{-.05in}}=\matrix{cc}{0.5\alpha_1 & 0.1 \\ 0.3 & 0.5\alpha_2}\matrix{c}{\hspace{-.05in}x_1(k)\hspace{-.05in}\\\hspace{-.05in}x_2(k)\hspace{-.05in}}+ \matrix{c}{1\\0}u(k),
$$
and
$$
y(k)=\matrix{cc}{1& 0}\matrix{c}{x_1(k)\\x_2(k)},
$$
where $x(k)\in\mathbb{R}^2$ and $u(k)\in\mathbb{R}$ are the state vector and the control input, respectively. Let us define the parameter vector as $\alpha=\matrix{cc}{\alpha_1 & \alpha_2}^\top\in\mathcal{A}\subset \mathbb{R}^2$ with
$$
\mathcal{A}=\left\{\alpha\in\mathbb{R}^2 | \;q_1(\alpha)=1-\alpha_1^2\geq 0,q_2(\alpha)=1-\alpha_2^2\geq 0 \right\}.
$$
We are interested in reducing the system complexity by getting a new model which is only a function of $\alpha_1$. First, let us present the model reduction algorithm introduced in~\cite{beck1996model}. To do so, we need to introduce the following notations
$$
A(\alpha)=A_0+\alpha_1A_1U_1+\alpha_2A_2U_2,
$$
where
$$
A_0=\matrix{cc}{0 & 0.1 \\ 0.3 & 0}\hspace{-.04in},\;A_1=\matrix{cc}{0.5 & 0 }\hspace{-.04in},\;
A_2=\matrix{cc}{0 & 0.5}\hspace{-.04in},
$$
and
$$
U_1=\matrix{cc}{1 & 0 }^\top,\;U_2=\matrix{cc}{0 & 1}^\top.
$$
Now, using~\cite{4048278Morton}, it is evident that
\begin{equation*}
\begin{split}
G(\mathfrak{z};\alpha)&=C(\mathfrak{z}I-A(\alpha))^{-1}B
=\left[\begin{array}{cccc} A_0 & U_1 & U_2 & B_0 \\ A_1 & 0 & 0 & 0 \\ A_2 & 0 & 0 & 0 \\ C_0 & 0 & 0 & 0 \end{array} \right]\hspace{-.04in}\star\hspace{-.04in} \left[\begin{array}{ccc} \hspace{-.1in}\mathfrak{z}^{-1}I_{2\times 2} \hspace{-.1in}& 0 & 0 \\ 0 & \alpha_1 & 0  \\ 0 & 0 & \alpha_2  \end{array} \right],
\end{split}
\end{equation*}
where $\star$ denotes the upper linear fractional transformation operator (see~\cite{beck1996model,4048278Morton} for its definition). Let us introduce notations
$$
\bar{A}=\matrix{ccc}{ A_0 & U_1 & U_2 \\ A_1 & 0 & 0 \\ A_2 & 0 & 0},\;
\bar{B}=\matrix{c}{ B_0 \\ 0 \\  0},\;
\bar{C}=\matrix{c}{C_0^\top \\ 0 \\ 0 }^\top.
$$
To get the reduced system, we need to solve the optimization problem
\begin{equation} \label{eqn:optimization3}
\begin{array}{rl}  \min_{X,Y\in \mathcal{W}} &  \trace(XY), \\ \mbox{subject to} & \bar{A}^\top X \bar{A}-X+\bar{C}^\top\bar{C} \leq 0, \\ & \bar{A} Y \bar{A}^\top -Y+\bar{B}\bar{B}^\top \leq 0,
\end{array}
\end{equation}
where
\begin{equation*}
\begin{split}
\mathcal{W}=\big\{W\in\mathcal{S}_+^{4}\;|\; W=\diag(W_{11},W_{22},W_{33}) \mbox{ such that } W_{11}\in\mathcal{S}_+^{2},W_{22}\in\mathcal{S}_+^{1},W_{33}\in\mathcal{S}_+^{1}, \big\}.
\end{split}
\end{equation*}
We use Procedure~\ref{alg:2} for solving the optimization problem in~(\ref{eqn:optimization3}). Using~\cite{618250}, we know that if the procedure is initialized correctly (i.e., close enough to the optimal solution), the algorithm converges to the optimal decision variables. Now, using matrices $X,Y\in\mathcal{W}$, we introduce the change of variable $T=\diag(T_0,T_1,T_2)$ to get the balanced realization of the system
\begin{equation*}
\begin{split}
T^{-1}\bar{A}T=\matrix{ccc}{T_0^{-1}A_0T_0 & T_0^{-1}U_1T_1 & T_0^{-1}U_2T_2 \\ T_1^{-1}A_1T_0 & 0 & 0 \\ T_2^{-1}A_2T_0 & 0 & 0},\;T^{-1}\bar{B}=\matrix{c}{ T_0^{-1}B_0 \\ 0 \\  0},\;
\bar{C}T=\matrix{ccc}{C_0T_0 & 0 & 0 }.
\end{split}
\end{equation*}
 Let us for the moment focus on just removing parameter~$\alpha_2$ from the model matrices (and not decreasing the order of the system). Using balanced truncation, we can calculate the reduced system as $G_r(\mathfrak{z};\alpha_1)=C_r(\mathfrak{z}I-A_r(\alpha_1))^{-1}B_r,$
where
\begin{equation*}
\begin{split}
A_r(\alpha_1)&=T_0^{-1}A_0T_0 +\alpha_1 T_1^{-1}A_1U_1T_1=\matrix{cc}{0.5\alpha_1 & -1.7\times 10^{-1} \\ -1.7\times 10^{-1} & 0},
\end{split}
\end{equation*}
$$
B_r=T_0^{-1}B_0=\matrix{cc}{-1.0 & 0.0}^\top, \; C_r=C_0T_0=\matrix{cc}{-1.0 & 0.0}.
$$
Finally, we can calculate the error caused by the parameter reduction as
\begin{equation*}
\begin{split}
\max_{\alpha\in\mathcal{A}}\|G_r(\mathfrak{z};\alpha_1)-G(\mathfrak{z};\alpha)\|_\infty=
0.14 \leq 2\sqrt{\sigma(X_{33}Y_{33})}=0.62,
\end{split}
\end{equation*}
where the upper bound of this error was introduced in~\cite{beck1996model}.

\begin{algorithm}[t!]
\caption{}
\label{alg:2}
\begin{algorithmic}
\REQUIRE Threshold $\epsilon\in\mathbb{R}_{>0}$.
\ENSURE $X,Y\in\mathcal{W}$.
\STATE \hspace{-.13in}\textbf{Initialization:} $X_{\mathrm{old}},Y_{\mathrm{old}}\in\mathcal{W}$.
\FOR{$k=1,2,\dots$}
\STATE Solve the optimization problem
$$
\begin{array}{rl}  \min_{X,Y\in \mathcal{W}} & \trace(X_{\mathrm{old}}Y+XY_{\mathrm{old}}), \\ \mbox{subject to} & \bar{A}^\top X \bar{A}-X+\bar{C}^\top\bar{C} \leq 0, \\ & \bar{A} Y \bar{A}^\top -Y+\bar{B}\bar{B}^\top \leq 0,
\end{array}
$$
\IF {$\|X-X_{\mathrm{old}}\|+\|Y-Y_{\mathrm{old}}\|\leq \epsilon$}
\STATE \textbf{break}
\ENDIF
\STATE $X_{\mathrm{old}} \leftarrow X$.
\STATE $Y_{\mathrm{old}} \leftarrow Y$.
\ENDFOR
\end{algorithmic}
\end{algorithm}

Now, we can illustrate the result of our proposed algorithm on this numerical example. Let us fix the polynomial degrees $d_A=1$, $d_B=1$, $d_C=0$, $d_D=0$, $d_P=2$, $d_{Q_0}=2$, $d_{Q_1}=0$, and $d_{Q_2}=0$. We use Procedure~\ref{alg:1} when adapted for discrete-time systems to get the optimal reduced system with $n'=2$. The resulting reduced system is
\begin{equation*}
\begin{split}
A'(\alpha_1)=\matrix{cc}{5.0\times 10^{-1}\alpha_1 & -1.2\times 10^{-1} \\ -3.3\times 10^{-1} & -6.5\times 10^{-4}\alpha_1},B'(\alpha_1)=\matrix{cc}{ 1.0 & 6.3\times 10^{-2}\alpha_1}^\top,
C'=\matrix{cc}{1.0 & 0 },\; D'=7.9\times 10^{-3}.
\end{split}
\end{equation*}
For this reduced system, we have
$$
\max_{\alpha\in\mathcal{A}}\|G'(\mathfrak{z};\alpha_1)-G(\mathfrak{z};\alpha)\|_\infty =0.095.
$$
As we can see, for this particular example, we could achieve a smaller distance between the transfer functions of the original system and its reduced one.

We can also try to reduce the system order by choosing $n'=1$. We use Procedure~\ref{alg:1} when adapted for discrete-time systems to get the optimal reduced system as
\begin{equation*}
\begin{split}
A'(\alpha_1)=5.2\times 10^{-1}\alpha_1 - 2.0\times 10^{-8},\;
B'(\alpha_1)=9.4\times 10^{-9} \alpha_1 + 9.9\times 10^{-1},\;
C'=1.0,\; D'=1.8\times 10^{-8}.
\end{split}
\end{equation*}
For this reduced system, we have
$$
\max_{\alpha\in\mathcal{A}}\|G'(\mathfrak{z};\alpha_1)-G(\mathfrak{z};\alpha)\|_\infty =0.19,
$$
 while if where using the method in~\cite{beck1996model}, we would have recovered
$$
\max_{\alpha\in\mathcal{A}}\|G_r(\mathfrak{z};\alpha_1)-G(\mathfrak{z};\alpha)\|_\infty =0.27,
$$
where $G_r(\mathfrak{z};\alpha_1)=C_r(\mathfrak{z}I-A_r(\alpha_1))^{-1}B_r$ with
\begin{equation*}
\begin{split}
A_r(\alpha_1)=5.0\times 10^{-1}\alpha_1,\;
B_r(\alpha_1)=-1.0,\;C_r=-1.0.
\end{split}
\end{equation*}

\subsection{Continuous-Time Systems}
In this subsection, we present a practical continuous-time numerical example. Let us consider a simple power network composed of two generators (see Figure~\ref{figure1}). We have partially extracted the structure of this example and its nominal numerical values from~\cite{Ghandhari00}. We can model this power network as
\begin{equation*}
\begin{split}
\dot{\delta}_1(t)&=\omega_1(t), \\
\dot{\omega}_1(t)&=\frac{1}{M_1}\big[P_{1}(t)-x_{12}^{-1}\sin(\delta_1(t)-\delta_2(t)) -x_{1}^{-1}\sin(\delta_1(t))-D_1\omega_1(t) \big],
\end{split}
\end{equation*}
and
\begin{equation*}
\begin{split}
\dot{\delta}_2(t)&=\omega_2(t), \\
\dot{\omega}_2(t)&=\frac{1}{M_2}\big[P_{2}(t)-x_{12}^{-1}\sin(\delta_2(t)-\delta_1(t)) -x_{2}^{-1}\sin(\delta_2(t))-D_2\omega_2(t) \big],
\end{split}
\end{equation*}
where $\delta_i(t)$ is the phase angel of the terminal voltage of the generator~$i$, $\omega_i(t)$ is its rotation frequency, and $P_{i}(t)$ is mechanical input power to the generator. We assume that $P_1(t)=1.6+u_1(t)$ and $P_2(t)=1.2+u_2(t)$, where $u_1(t)$ and $u_2(t)$ are the control inputs to the system. The power network parameters can be found in Table~\ref{table:1} (see~\cite{Ghandhari00} for a description of these parameters). Note that all these values are given in per unit. Now, we can find the equilibrium point $(\delta_1^*,\delta_2^*)$ of these nonlinear coupled systems and linearized the overall system around its equilibrium which would result in
\begin{equation} \label{eqn:longequation}
\begin{split}
\frac{\mathrm{d}}{\mathrm{d}t} \hspace{-.04in}\matrix{c}{ \hspace{-.06in} \Delta\delta_1(t) \hspace{-.06in} \\ \hspace{-.06in} \Delta\omega_1(t) \hspace{-.06in} \\ \hspace{-.06in} \Delta\delta_2(t) \hspace{-.06in} \\ \hspace{-.06in} \Delta\omega_2(t) \hspace{-.06in}} \hspace{-.06in}=\hspace{-.06in}\matrix{cccc}{0 & 1 & 0 & 0 \\ \frac{-x_{12}^{-1}\cos(\delta_1^*-\delta_2^*)-x_1^{-1}\cos(\delta_1^*)}{M_1} & -\frac{D_1}{M_1} & \frac{\cos(\delta_1^*-\delta_2^*)}{x_{12}M_1} & 0 \\ 0 & 0 & 0 & 1 \\ \frac{\cos(\delta_2^*-\delta_1^*)}{x_{12}M_2} & 0 & \frac{-x_{12}^{-1}\cos(\delta_2^*-\delta_1^*)-x_2^{-1}\cos(\delta_2^*)}{M_2} & -\frac{D_2}{M_2}}
\matrix{c}{ \hspace{-.06in} \Delta\delta_1(t) \hspace{-.06in} \\ \hspace{-.06in} \Delta\omega_1(t) \hspace{-.06in} \\ \hspace{-.06in} \Delta\delta_2(t) \hspace{-.06in} \\ \hspace{-.06in} \Delta\omega_2(t) \hspace{-.06in}}
\hspace{-.06in}+\hspace{-.06in}\matrix{cc}{0 & 0 \\1 & 0 \\0 & 0 \\0 & 1}\hspace{-.06in}\matrix{c}{\hspace{-.06in}u_1(t)\hspace{-.06in} \\ \hspace{-.06in}u_2(t)\hspace{-.06in}}\hspace{-.04in},
\end{split}
\end{equation}
where $\Delta\delta_1(t)$, $\Delta\delta_2(t)$, $\Delta\omega_1(t)$, and $\Delta\omega_2(t)$ denote the deviation of the state variables $\delta_1(t)$, $\delta_2(t)$, $\omega_1(t)$, and $\omega_2(t)$ from their equilibrium points. Let us assume that we have connected impedance loads to each generator locally. Hence, the parameters $x_1$ and $x_2$ can vary over time according to the load profiles. Furthermore, assume that each generator changes its input mechanical power according these local load variations. Doing so, we would not change the equilibrium point $(\delta_1^*,\delta_2^*)$. For this setup, we can model the system as a continuous-time parameter-dependent linear system described by
\begin{equation*}
\begin{split}
\begin{array}{rl} G(s;\alpha): & \hspace{-.1in}\left\{\hspace{-.06in}\begin{array}{l} \dot{x}(t)=A(\alpha)x(t)+Bu(t),\\
y(t)=Cx(t)+Du(t), \end{array} \right. \end{array}
\end{split}
\end{equation*}
where
\begin{equation} \label{eqn:longequation1}
\begin{split}
A(\alpha)=\matrix{cccc}{ 0 & 1 & 0 & 0 \\ -2.7\times 10^{1}\alpha_1-1.5\times 10^{2} & -2.5\times 10^{-1} & 9.8\times 10^{1} & 0 \\ 0 & 0 & 0 & 1 \\ 7.8\times 10^{1} & 0 & -2.3\times 10^{1}\alpha_2- 1.2\times 10^{2} & -2.0 \times 10^{-1} }\hspace{-.05in}.
\end{split}
\end{equation}
and
$$
B=\matrix{cccc}{0 &1 &0 &0 }^\top\hspace{-.05in},\hspace{.2in} C=\matrix{cccc}{1&0&0&0}\hspace{-.07in}, \hspace{.2in} D=0,
$$
with $\alpha=[\alpha_1\;\alpha_2]^\top$. In this formulation, parameter $\alpha_i$ for $i=1,2$, denotes the deviation of the admittance $x_i^{-1}$ from its nominal value (see Table~\ref{table:1}). Note that here we have chosen the input-output pair to derive a reduced model for the network from the perspective of the first generator. One can try to solve this problem for any other given set of inputs and outputs. We assume that
$$
\mathcal{A}=\left\{\alpha\in\mathbb{R}^2\;|\;0.1^2-\alpha_i^2\geq 0 \mbox{ for } i=1,2\right\}.
$$
Let us fix the polynomial degrees $d_A=1$, $d_B=0$, $d_C=0$, $d_D=0$, $d_P=3$, $d_{Q_0}=2$, $d_{Q_1}=2$, and $d_{Q_2}=2$. We use Procedure~\ref{alg:1} to get the optimal reduced system with $n'=4$. The resulting reduced system is
\begin{equation*}
\begin{split}
\begin{array}{rl} G'(s;\alpha_1): & \hspace{-.1in}\left\{\hspace{-.06in}\begin{array}{l} \dot{x}(t)=A'(\alpha_1)x(t)+B'u(t),\\
y(t)=C'x(t)+D'u(t), \end{array} \right. \end{array}
\end{split}
\end{equation*}
where 
\begin{small}
\begin{equation} \label{eqn:longequation2}
\begin{split}
A'(\alpha_1)\hspace{-.01in}=\hspace{-.07in}\matrix{cccc}{ \hspace{-.07in} 7.4\times 10^{-4}\alpha_1 - 4.2 & \hspace{-.07in} 5.5\times 10^{-4}\alpha_1 + 2.6\times 10^{-2} & \hspace{-.07in}1.7\times 10^{-4}\alpha_1 + 8.8\times 10^{-3} & \hspace{-.07in}-7.5\times 10^{-4}\alpha_1 + 7.3\times 10^{-2} \hspace{-.07in}\\  \hspace{-.07in}-3.0\times 10^{1}\alpha_1 + 4.4\times 10^{-1} & \hspace{-.07in} 1.9\times 10^{-1}\alpha_1 - 6.4 & \hspace{-.07in}-4.4\times 10^{-2}\alpha_1 - 2.8\times 10^{-1} & \hspace{-.07in}-1.4\times 10^{-1}\alpha_1 - 4.4\times 10^{-1} \hspace{-.07in}\\  \hspace{-.07in} -6.6\times 10^{-3}\alpha_1 + 1.5 & \hspace{-.07in} -2.3\times 10^{-4}\alpha_1 + 3.8\times 10^{-2} & \hspace{-.07in} 2.8\times 10^{-4}\alpha_1 - 4.2\times 10^{-1} & \hspace{-.07in} -1.0\times 10^{-4}\alpha_1 + 4.6\times 10^{-2} \hspace{-.07in}\\ \hspace{-.07in} 5.9\times 10^{-2}\alpha_1 - 1.8\times 10^{-2} & \hspace{-.07in} -4.7\times 10^{-5}\alpha_1 - 8.8\times 10^{-1} &  \hspace{-.07in} 4.2\times 10^{-3}\alpha_1 + 1.8\times 10^{-1} & \hspace{-.07in} -5.5\times 10^{-3}\alpha_1 - 8.3\times 10^{-1} \hspace{-.07in}}\hspace{-.05in},
\end{split}
\end{equation}
\end{small}
and
$$
B'=\matrix{cc}{9.0\times 10^{-2}\\3.8\\-4.9\times 10^{-2}\\5.6\times 10^{-1}}\hspace{-.05in},\hspace{.05in} C'=\matrix{c}{1.1\times 10^{-1} \\ 4.3\times 10^{-1} \\ -3.2 \times 10^{-2}\\ 1.1\times 10^{-1}}^\top\hspace{-.07in},\;D'=-0.10558.
$$
For this reduced system, we have
$$
\max_{\alpha\in\mathcal{A}}\|G'(s;\alpha_1)-G(s;\alpha)\|_\infty =1.5\times 10^{-1}.
$$
Hence, we could effectively remove the model matrices dependencies on the second subsystem parameter $\alpha_2$ while not drastically changing the first subsystem input-output transfer function.

\jpfig{0.35}{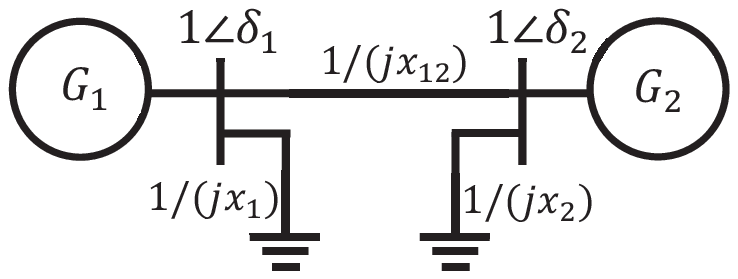}{figure1}{Schematic diagram of the power network in our numerical example.}

\renewcommand{\arraystretch}{1.3}
\begin{table}
\caption{Nominal value of power system parameters extracted from~\cite{Ghandhari00}. }
\label{table:1}
\centering
\begin{tabular}{|c||c|} \hline
Variable &  Nominal Value (p.u.) \\ \hline \hline
$M_1$ &  $2.6 \times 10^{-2}$ \\ \hline
$M_2$ &  $3.2 \times 10^{-2}$ \\ \hline
$x_{12}$ & $4.0\times 10^{-1}$ \\ \hline
$x_1$ & $5.0 \times 10^{-1}$ \\ \hline
$x_2$ & $5.0 \times 10^{-1}$ \\ \hline
$D_1$ & $6.4\times 10^{-3}$ \\ \hline
$D_2$ & $6.4\times 10^{-3}$ \\ \hline
\end{tabular}
\end{table}

\section{Conclusions and Future Work} \label{sec:conclusion}
In this paper, we presented a powerful procedure for approximating parameter-dependent linear systems with less complex ones using fewer model parameters or state variables. To do so, we minimized the distance between the transfer function of the original system and its reduced version. We presented a suboptimal method for solving this minimization problem using sum-of-squares optimization and the method of alternating LMIs for solving BMIs. We developed numerical procedures for both continuous-time and discrete-time system contrary to the available result which focused mostly on discrete-time systems.
Due to relying on sum-of-square optimization, the developed procedures would not scale well with the system dimension and the number of parameters. Possible future research could focus on developing a better numerical procedure for dealing with BMIs and studying the convergence properties of this numerical approach. Furthermore, we could exploit sparsity patterns or symmetry structures to improve the scalability of the sum-of-square optimization.

\bibliography{ref}
\bibliographystyle{ieeetr}

\end{document}